\documentclass{article}
\usepackage{amsfonts,amssymb,amsmath}
\usepackage[dvips]{epsfig}
\usepackage[T1]{fontenc}
\usepackage[latin1]{inputenc}
\usepackage{graphicx}
\usepackage[english]{babel}
\usepackage{amsmath}
\usepackage{amssymb}
\usepackage{amsfonts}
\usepackage{color}

\begin{document}

\title{\bf  Deformed Ho\v{r}ava-Lifshitz Cosmology\\
 and \\ Stability of   Einstein Static
Universe }
\author{Y. Heydarzade$^{1}$\thanks{%
email: heydarzade@azaruniv.edu}\,, M. Khodadi $^{2}$\thanks{%
email: m.khodadi@iaufb.ac.ir} and F. Darabi$^{1,3}$\thanks{%
email: f.darabi@azaruniv.edu; Corresponding author}
\\{\small $^1$ Department of Physics, Azarbaijan Shahid Madani University, Tabriz, 53714-161, Iran}\\$^2${\small Young Researchers and Elite Club, Firoozkooh Branch, Islamic Azad University, Firoozkooh, Iran }\\ {\small $^3$Research Institute for Astronomy and Astrophysics of Maragha (RIAAM), Maragha 55134-441, Iran}}
\date{\today}

\maketitle

\begin{abstract}
Stability of the Einstein static universe versus the linear scalar, vector
and tensor perturbations is  investigated in the context of  deformed Ho\v{r}ava-Lifshitz cosmology inspired by entropic force scenario. A general stability condition against the linear scalar perturbations is obtained. Using this general condition, it is shown that there is no stable Einstein static universe for the case of flat universe, $k=0$. For the the special case of large values of running parameter of HL gravity $\omega$, in a positively curved universe $k>0$, the domination of the quintessence and phantom matter fields with barotropic equation of state parameter $\beta<-\frac{1}{3}$ is necessary while for a negatively curved universe $k<0$, the matter fields with $\beta> -\frac{1}{3}$ are needed to be the dominant fields of the universe. Also, a neutral stability
against the vector perturbations is obtained. Finally, an inequality including the cosmological parameters of the Einstein static universe is obtained for the stability against the tensor perturbations. It turns out that for large $\omega$ values, there is a stability against the tensor perturbations.
\\
\\
Keywords: Ho\v{r}ava-Lifshitz cosmology, Einstein static universe, stability
\end{abstract}

\section{Introduction}
While near a century has passed since the birth of General Relativity (GR),
we still do not have a clear understanding of the origin of the gravitational
force. In the quest for discovering the nature of this mysterious force, we are
faced with two significant problems. The first is the lack of a theoretical
framework to reconcile  two prosperous theories; the quantum mechanics and GR
at the Planck scale. The second is the absence of a unified theory of gravitational
force with other three fundamental forces. Contrary to the common approaches for
solving these problems, E. Verlinde \cite{Verlinde} by continuing the works done by
Sakharov \cite{Sakharov}, Jacobson \cite{Jacobson} and Padmanabhan \cite{Padmanabhan},
 claimed that the gravity is not a fundamental but is an emerging force in
the spacetime. Indeed, gravity can be described as an entropic force due to the changes
of information on the holographic screen when a test particle situated at an arbitrary
distance from the screen  is shifting toward it. Generally, the approach of Verlinde is
built on two main pillars; the holographic principle and the equipartition rule which
reproduces Newton's laws and Einstein's field equations. Undoubtedly, if the validity of
the current interpretation of gravity be approved, then it will affect the direction of
research in theoretical physics at the future.  As an example,
the author of \cite{Lee} by using the Verlinde's hypothesis, extracted the Newtonian gravity
in  loop quantum gravity. It is also interesting that in \cite{Wang}, the entropic force is
considered as the origin of the coulomb force. In \cite{Caravelli}, it is shown that the
holographic dark energy will be derivable from the entropic force interpretation of gravity.
The modified Friedmann equations by this notion of gravity are obtained in different setups
such as Einstein gravity \cite{Einstein}, braneworld scenarios \cite{Brane}, Gauss-Bonnet
gravity \cite{Ahmad}, and  Ho\v{r}ava-Lifshitz gravity
\cite{Liu}. There are also  other applications of Verlinde's hypothesis,
some of which are mentioned  in \cite{other}.

As mentioned above, theoretical physics
suffers from the lack of a quantum theory of gravity. As one of the biggest hurdle in achieving
this goal, one can  point to the fact that GR is not renormalizable theory at  high energy limits
(ultra-violet limit (UV)), so there is no control on the theory and its predictions. As one of
the efforts made to overcome  this big problem, one can  refer to a new class of UV complete theory
of gravity known as Horava-Lifshitz (HL) gravity \cite{Horava}. We should note that  HL theory of
gravity is achieved at the cost of loosing the Lorentz symmetry through  a Lifshitz-type anisotropic
scaling at  high energy limits i.e. $t\rightarrow l^{z}t, \quad x^{i}\rightarrow lx^{i}$ where $z\geq1$
is the dynamical critical exponent. Therefore, it is a non-relativistic renormalizable
theory of gravity.  While  HL gravity at  high energies is non-
relativistic, it is expected that the four dimensional general covariance can be recovered at the low
energy limits. More technically, in contrast to the standard GR, the HL theory of gravity is not full
diffeomorphism invariance, rather it only has a local Galilean invariance \cite{Masato}. Then, it is expected
that HL gravity approaches to GR at the infra-red (IR) limits. However, the HL theory of gravity can be considered
as a UV complete candidate for GR. The existence of an anisotropic scaling at the UV limit results in a
mechanism for generation of cosmological perturbations so that it can solve the horizon problem without resorting
to inflation. It is also a remarkable point that in  HL model of gravity, due to the lack of local Hamiltonian
constraint, dark matter may appear as an integration constant \cite{Mukohyama}. In order to a comprehensive review
of HL gravity and some of its cosmological implications,  see \cite{S}. The total action of HL
gravity can be written as
\cite{Horava}
\begin{eqnarray}\label{e1-1}
S_{HL}& = &\int dt dx^{3} \sqrt{g}N \left[\frac{2}{\kappa^{2}}(K_{ij}K^{ij}-\lambda K^{2})-\frac{\kappa^{2}}{\omega^{4}}
C_{ij}C^{ij}+\frac{\kappa^{2}\mu}{2\omega^{2}} \epsilon^{ijk}\, {^3}\!R_{il}\nabla_{j}\,{^3}\!R_{k}^{l}
\right. \nonumber\\ & &\left.-\frac{\kappa^{2}\mu^{2}}{8}\,{^3}\!R_{il}\,{^3}R^{ij}
 + \frac{\kappa^{2}\mu^{2}}{8(3\lambda-1)}\left(\frac{4\lambda-1}{4}\,({^3}\!R)^{2}-
\Lambda\,{^3}\!R+3\Lambda^{2} \right)\!R\right]\;,
\end{eqnarray}
where $ \mu, \Lambda$ and $\kappa, \lambda, \omega$ represent the dimensional and dimensionless constant
parameters of HL gravity, respectively. Also, the quantities $K_{ij}=\frac{1}{2N}(\dot{g}-
\nabla_{i}N_{j}-\nabla_{i}N_{j})$ and $C^{ij}=\epsilon^{ijk}\nabla_{k}(^{3}\!R_{l}^{j}-\frac{1}{4}\,
^{3}\!R\,\delta^{i}_{j})$ denote the extrinsic curvature and the Cotton tensor, respectively.
Note that for the case of $\lambda\rightarrow1$, the kinetic section of action (\ref{e1-1})
 approaches to GR action in IR limit. It should be stressed that in UV limit, the behavior of HL model of
gravity is very different from GR. So, as will be discussed in the following,  for  $\lambda\rightarrow1$
  HL gravity does not perfectly reproduces full four dimensional diffeomorphism invariance at large distances
(IR limit).

One of the most interesting issues in the context of any model of gravity is the black hole
solutions. For the first time,  the authors of \cite{Pop} provided a static spherically symmetric black hole solution
for asymptotically Lifshitz spacetimes in the presence of running coupling constant $\lambda$. Through the study of
thermodynamical properties of this solution, it is shown in \cite{Castillo} that for the case of $\lambda=1$, this solution
reduces to Reissner-Nordstrom black hole instead of the standard Schwarzschild black hole. More precisely, since the
horizon radius of the black hole solution, corresponding to $\lambda=1$, includes a geometric parameter as $\alpha=\frac{1}{2\omega}$
which can play the role of electric charge, it looks like the Reissner-Nordstrom black hole. While the geometric parameter $\alpha$
results in the modified entropy of the black hole, for increasing $\omega$ it becomes smaller and in the limit $\omega\rightarrow\infty$
it approaches to zero and one recovers the standard Schwarzschild entropy expression, $S=\frac{A}{4}$. With these properties, the original
action in equation (\ref{e1-1}) represents the \emph{deformed} HL model of gravity in which one recovers
the standard GR in IR limit for case $\lambda=1$,  while its black hole solution does not result in reproduction of the usual Schwarzschild black hole solution.
Of course, by adding an IR modification term $\mu^{4}$ $^{(3)}\!R$ to the
original HL action (\ref{e1-1}) this issue will be resolved
(see \cite{park} for more details). Also, it should be mentioned that
the action (\ref{e1-1}) is not the most general action of the HL gravity. In Ref.\cite{PRL},
a natural extension of the original version of HL theory, regarding the terms depending on the acceleration
of the foliation, has been presented. This extended version is power-counting renormalizable and free of notorious pathologies appeared in the original HL model \cite{Horava}. A main feature of
the extended version is that in the IR limit it reduces to a Lorentz-violating scalar-tensor gravity
theory instead of GR. However,  it is shown that this inconsistency with GR can be improved
by an appropriate selection of parameters of the theory.

In this work, we investigate the Einstein static universe and its stability versus homogeneous scalar,
vector and tensor perturbations in the framework of the deformed HL cosmology inspired by entropic force
scenario. The main motivation of studying the stability of Einstein static universe comes from the emergent
universe scenario \cite{Ellis}. The emergent universe scenario is a past-eternal inflationary model in which
the horizon problem is solved before the beginning of inflation and the big-bang singularity is removed. In
the framework of this cosmological model, the universe is originated from an Einstein static state rather than
a big bang singularity. However, this model suffers from a fine-tuning problem which can be ameliorated by modifying
the cosmological field equations of the general relativity. For this reason, analogous static solutions and their
stabilities have been studied in the context of the modified theories of gravity such as $f(R)$  \cite{f(R)}, $f(T)$
\cite{f(T)}, Einstein-Cartan theory \cite{Cartan}, massive gravity \cite{massive}, Lyra geometry \cite{Darabi},
loop quantum cosmology \cite{Loop} and braneworld scenarios \cite{brane}. Stability of the Einstein static universe
is also studied in the Horava-Lifshitz  gravity \cite{HL}. Our present paper is based on the model proposed in \cite{Liu}
in which the dynamical equations of the deformed HL gravity for a Friedmann-Robertson-Walker (FRW) background is obtained based
on the thermodynamical properties of the deformed HL black holes \cite{Castillo}. In the present framework of deformed HL cosmology
inspired by entropic gravity, the modified Friedmann equations and the corresponding  results  are different from those obtained
in Refs. \cite{HL}. Throughout this work, we use the units of $\hbar=c=G=k_{B}=1$.

\section{Friedmann-Robertson-Walker Universe in
Deformed HL Gravity from Entropic Force}
In this section, assuming the emergence of gravity in space-time as an entropic force,
we plan to derive the modified Friedmann equations in HL cosmological setup. We consider
the homogeneous and isotropic FRW background space-time with the metric
\begin{equation}\label{e2-1}
ds^{2}=h_{ab}dx^{a}dx^{b}+\tilde{r}^{2}d\Omega_{2}^{2}\;,
\end{equation}
where $h_{ab}= \mbox{diag}\left(-1,\frac{a^{2}}{1-kr^{2}}\right)$ and $d\Omega_{2}^{2}=
d\theta^{2}+\sin^{2}d\phi^{2}$  denote the two dimensional metrics, $\tilde{r}= a(t)r$
in which $a(t)$ is the cosmic scale factor and $k=-1,\,0$ or $1$ corresponds to an open, flat
or closed universe, respectively. In a FRW spacetime,  one can show that
\begin{equation}\label{e2-2}
\tilde{r}_{A}=\frac{1}{\sqrt{H^{2}+\frac{k}{a^{2}}}}\;,
\end{equation}
where $\tilde{r}_{A}$ and $H=\frac{\dot{a}}{a}$, are the dynamical apparent horizon and
the Hubble parameter, respectively. In order to {have a comprehensive review on trap
surfaces and dynamical apparent horizons, see \cite{Hayward}. The temperature on the
apparent horizon reads as $T=\frac{|\kappa_{sg}|}{2\pi}$  where $\kappa_{sg}$ is the surface
gravity whose value at the
apparent horizon of the FRW universe is given by
\begin{equation}\label{e2-3}
\kappa_{sg}=-\frac{1}{\tilde{r}_{A}}\left(1-\frac{\dot{\tilde{r}}_{A}}{2H\tilde{r}_{A}}\right).
\end{equation}
By assuming that the universe as a closed system is in the thermodynamical equilibrium,
one may ignore the second term in equation (\ref{e2-3}) and find the following
temperature of the apparent horizon as
\begin{equation}\label{e2-4}
T_{A}=\frac{1}{2\pi \tilde{r}_{A}}.
\end{equation}
In Verlinde's scenario of gravity, the entropy of black hole has a very important role in the derivation of Newton's law of
gravitation and Friedmann equation, such that in the presence of any correction, entropy-area relation in Einstein
gravity is extended as
\begin{equation}\label{e2-5}
S=\frac{A}{4}+s(A)\;,
\end{equation}
where $A$ and $s(A)$ are the area of horizon and entropy correction term, respectively.
It should be noted that the Loop quantum gravity (LQG)
imposes a quantum correction to the
horizon area law of a black hole, namely the logarithmic correction, as follows \cite{Cai}
\begin{equation}\label{e2-5a}
S=\frac{A}{4}+\frac{\alpha}{4}\ln({A})\;,
\end{equation}
where $\alpha$  is a dimensionless constant  with order of unity.
On the other hand, in Ref.\cite{Castillo} it has been shown that the entropy of black holes in the
context of  HL gravity and in the presence of logarithmic correction, can be written as
\begin{equation}\label{e2-6}
S=\frac{A}{4}+\frac{\pi}{\omega}\ln({A}),
\end{equation}
where $\omega$ refers to the dimensionless running coupling constant of HL gravity.
In order to have a black hole within the framework of deformed HL theory,  the
condition $M^{2}\geq(2\omega)^{-1}$ must be satisfied where $M$ denotes an integration constant which can
be interpreted as the mass of black hole as explained in \cite{Castillo}.
The equality case corresponds to the extremal 
black hole with a horizon area as $M^{2}=r_{e}^{2}=(2\omega)^{-1}$. By
comparing the logarithmic correction terms in equations (\ref{e2-5a}) and (\ref{e2-6}), one finds that  $\alpha=4\pi\omega^{-1}$. As a result, the
dimensionless constant
$\alpha,$ coming from LQG, gets a geometric interpretation
and is known as \emph{``geometric parameter''}.
Equations (\ref{e2-6}) and (\ref{e2-5a}) indicate that for the special case of
$\omega\rightarrow\infty$ (or $\alpha\rightarrow0$) one will recover
the standard GR entropy expression. In this process,
by using the continuity equation
\begin{equation}\label{e2-7}
\dot{\rho}+3H(\rho+p)=0\;,
\end{equation}
and the energy equipartition rule
\begin{equation}\label{e2-8}
E =\frac{ 1}{2} NT\;,
\end{equation}
the Raychaudhuri equation can be expressed as
\begin{equation}\label{e2-9}
\frac{ 1}{2} NdT+ \frac{1}{2} T dN=4\pi\tilde{r}^{3}_{A}(\rho+p)H dt\;,
 \end{equation}
where $N$ denotes the number of bits on the screen and is proportional to
the area of the screen (horizon),  because of  $N=4S$.
As has already been mentioned, the approach of Verlinde is strongly influenced by the holographic principle
and equipartition rule. Therefore, inspired by holographic principle and assuming that the screen
has a total energy $E$, which is distributed between all the bits $N$, the temperature of the screen $T$ is
given by the equipartition rule in equation (\ref{e2-8}). Integrating
Raychaudhuri equation (\ref{e2-9}) will gives the following modified Friedmann
equation in the deformed HL cosmological setup inspired  by the entropic force \cite{Liu}}
\begin{equation}\label{e2-10}
\frac{1}{2\omega}(H^{2}+\frac{k}{a^2})^{2}+H^{2}+\frac{k}{a^2}=\frac{8\pi}{3}\rho.
\end{equation}
As we see, the Friedmann equation of the standard model of cosmology
is modified by the presence of the first term. As the simplest case, i.e. for the flat universe $k=0$, the
Friedmann equation is modified by a $\frac{H^4}{2\omega}$ term. Also, by differentiation of equation (\ref{e2-10})
with respect to the cosmic time  and using the continuity equation (\ref{e2-7}) the  following modified
acceleration equation can be obtained
\begin{equation}\label{e2-11}
\frac{\ddot a}{a}=\omega\left[-1+(1+\frac{16\pi}{3\omega}\rho)^{\frac{1}{2}}\right]
-4\pi\left[1+\frac{16\pi}{3\omega}\rho\right]^{-\frac{1}{2}}(\rho+p).
 \end{equation}
It is  easy to check that equations (\ref{e2-10}) and (\ref{e2-11}) in IR limit,
$\omega\rightarrow\infty$, will reduce to the following standard Friedmann equations as
\begin{equation}\label{e2-12}
H^{2}+\frac{k}{a^{2}}=\frac{8 \pi }{3}\rho.
\end{equation}
and
\begin{equation}\label{e2-13}
\frac{\ddot a}{a}= -\frac{4\pi}{3}(\rho+3p).
\end{equation}
 \section{Einstein Static Universe, Scalar Perturbations and Stability Analysis}
In what follows, we will consider the barotropic equation of state $p(t)=\beta\rho(t)$
and will expand the acceleration equation (\ref{e2-11}),  keeping up to the third order of the energy density
$\rho(t)$. The Einstein static universe in the deformed HL cosmological setup inspired by entropic gravity can
be obtained by the condition $\ddot a=\dot a=0$, through the equations (\ref{e2-10}) and (\ref{e2-11}) as
\begin{equation}\label{e3-1}
\frac{k^2}{2\omega a_0^4}+\frac{k}{a_0^2}=\frac{8\pi}{3}\rho_0\;,
\end{equation}
and
\begin{equation}\label{e3-2}
\omega\left[ \frac{8\pi}{3\omega}\rho_{0}-\frac{1}{8}(\frac{16\pi}{3\omega}\rho_{0})^2{+\frac{1}{16}(\frac{16\pi}{3\omega}\rho_{0})^3+}
...\right]-4\pi\left[1- \frac{8\pi}{3\omega}\rho_{0}+\frac{3}{8}(\frac{16\pi}{3\omega}\rho_{0})^2+
... \right]\rho_{0}(1+\beta)=0\;,
\end{equation}
where $a_0$ and $\rho_0$  refer to the scale factor and the energy density of the Einstein
static universe, respectively. We consider the linear homogeneous scalar perturbations around the
Einstein static universe, given in equations (\ref{e3-1}) and (\ref{e3-2}), and  investigate their
stability versus these perturbations. The perturbations in the cosmic scale factor $a(t)$ and the energy
density $\rho(t)$ depending only on time can be represented by
\begin{eqnarray}\label{e3-3}
&&a(t)\rightarrow a_{0}(1+\delta a(t)),\nonumber\\
&&\rho(t)\rightarrow \rho_{0}(1+\delta \rho(t)).
\end{eqnarray}
Substituting (\ref{e3-3}) into equation (\ref{e2-10}) and using  equation (\ref{e3-1}),
via linearizing the perturbation terms, leads to the following equation
\begin{equation}\label{e3-4}
-\left(\frac{k^2}{\omega a_0^4}+\frac{k}{a_0^2}\right)\delta a=\frac{4\pi}{3}\rho_0\delta\rho.
\end{equation}
By applying the same method on equations (\ref{e2-11}) and (\ref{e3-2}), we get
\begin{equation}\label{e3-5}
\ddot{\delta a}=\frac{4\pi}{3}\rho_0\delta\rho\left[ -1-3\beta+\frac{16\pi}{3\omega}(2+3\beta)\rho_{0}
-{\frac{32\pi^2}{3\omega^2}(7+9\beta)\rho_{0}^2}\right],
\end{equation}
where   substituting equation (\ref{e3-4}) into (\ref{e3-5}) gives the following equation
\begin{equation}\label{e3-6}
\ddot{\delta a}+\left(\frac{k^2}{\omega a_0^4}+\frac{k}{a_0^2}\right)\left[-1-3\beta+
\frac{16\pi}{3\omega}(2+3\beta)\rho_{0}-\frac{32\pi^2}{3\omega^2}(7+9\beta)\rho_{0}^2\right]\delta a=0.
 \end{equation}
Then, in order to have the oscillating perturbation modes representing the existence of a stable Einstein
static universe in the framework of the deformed HL cosmology from entropic gravity, the following condition
should be satisfied
\begin{equation}\label{e3-7}
\left(\frac{k^2}{\omega a_0^4}+\frac{k}{a_0^2}\right)\left[-1-3\beta+\frac{16\pi}{3\omega}(2+3\beta)\rho_{0}
-\frac{32\pi^2}{3\omega^2}(7+9\beta)\rho_{0}^2\right]>0,
\end{equation}
which leads to the following solution for the equation (\ref{e3-6})
\begin{equation}\label{e3-8}
\delta a=C_{1}e^{iAt}+C_{2}e^{-iAt},
\end{equation}
where $C_1$ and $C_2$ are integration constants and $A$ is given by
\begin{equation}\label{e3-9}
A=\left(\frac{k^2}{\omega a_0^4}+\frac{k}{a_0^2}\right)^{\frac{1}{2}}\left[-1-3\beta+\frac{16\pi}{3\omega}(2+3\beta)\rho_{0}
-\frac{32\pi^2}{3\omega^2}(7+9\beta)\rho_{0}^2\right]^{\frac{1}{2}}.
\end{equation}
The stability condition, together with the inequality (\ref{e3-7}), gives us the following different class of solutions:
\begin{itemize}
\item For the case of the flat universe, $k=0$, there is no stable Einstein static universe in the framework of
the deformed HL cosmological setup inspired by entropic gravity.
\item For  large $\omega$ values for which the first term in the parenthesis as well as  the third and
fourth terms in the brackets in equation (\ref{e3-7}) vanish, we will have the following condition
\begin{equation}
\frac{k}{a_0^2}(-1-3\beta)>0,
\end{equation}
which leads to the following cases:
\begin{enumerate}
\item The case of $k>0$ with the equation of state parameter $\beta<-\frac{1}{3}$,  representing the  quintessence
and phantom fields. \item The case of $k<0$ with the equation of state parameter $\beta>-\frac{1}{3}$, representing ordinary matter fields.
 \end{enumerate}
\item For a non-flat universe with an arbitrary $\omega$ values, the general constraint in equation (\ref{e3-7}) must be
satisfied which shows an interplay between the cosmological parameters $a_0$, $\rho_0$, $k$, $\omega$ and $\beta$ of this model.
\end{itemize}
\section{Vector and Tensor Perturbations and Stability Analysis}

In this section, we  study the stability analysis of the Einstein
static universe against the vector and tensor
perturbations. In a cosmological context, the vector perturbations
of a perfect fluid having energy density $\rho(t)$ with a barotropic pressure $p(t)=\beta
\rho(t)$ are governed by the co-moving dimensionless {\it vorticity} defined as ${\varpi}_a
=a{\varpi}$ whose modes satisfy the following propagation equation
\begin{equation}\label{e4-1}
\dot{\varpi}_{\kappa}+(1-3c_s^2)H{\varpi}_{\kappa}=0,
\end{equation}
where $c_s^2=dp/d\rho$ is the sound speed and $H$ is the Hubble parameter
\cite{tensor}. This equation is valid in our treatment of Einstein static universe in the
framework of the deformed HL gravity derived from entropic force scenario through the modified
Friedmann equations (\ref{e2-10}) and (\ref{e2-11}). For the Einstein static universe with
$H=0$, equation (\ref{e4-1}) reduces to
\begin{equation}\label{e4-2}
\dot{\varpi}_{\kappa}=0.
\end{equation}
Equation (\ref{e4-2}) represents that the initial vector perturbations will
remain frozen.
Then, independent of the  values of the the cosmological parameters $a_0$, $\rho_0$, $k$, $\omega$ and $\beta$ of this model,
there is a neutral stability against vector perturbations.

The tensor perturbations, namely gravitational-wave perturbations, of a perfect fluid are
described by the co-moving dimensionless transverse-traceless shear $\Sigma_{ab}=a\sigma_{ab}$,
whose modes satisfy the following equation
\begin{equation}\label{e4-3}
\ddot\Sigma_{\kappa}+3H\dot\Sigma_{\kappa}+\left[\frac{\mathcal{K}^2}{a^2}
+\frac{2k}{a^2}-\frac{8\pi}{3}(1+3\omega)\rho\right]\Sigma_{\kappa}=0,
\end{equation}
where $\mathcal{K}$ is the co-moving index $D^2\rightarrow -\mathcal{K}^2/a^2$ in which $D^2$ is
the covariant spatial Laplacian \cite{tensor}.  One can show that for the Einstein static universe,
this equation by using equations (\ref{e2-10}), (\ref{e2-11})
and (\ref{e3-1}), takes the form
\begin{equation}\label{e4-4}
\ddot\Sigma_{\kappa}+\left[\frac{1}{ a_0^2}(\mathcal{K}^{2}-\frac{k^2}{\omega
a_0 ^2})-2\omega\left(1-(1+\frac{16\pi}{3\omega}\rho_0
)^\frac{1}{2}\right)  \right]\Sigma_{\kappa}=0.
\end{equation}
Then, in order to have the stable modes against the tensor perturbations, the following inequality
should be satisfied
\begin{equation}\label{e4-5}
\frac{1}{ a_0^2}(\mathcal{K}^{2}-\frac{k^2}{\omega
a_0 ^2})-2\omega\left(1-(1+\frac{16\pi}{3\omega}\rho_0
)^\frac{1}{2}\right)>0.
\end{equation}
Using the expansion $(1+x)^{1/2}=1+\frac{1}{2}x-\frac{1}{8}x^{2}+O(x^3)$
\footnote{{Note that we have used the above expansion relation also in obtaining the equations (\ref{e3-2}) and (\ref{e3-5}).}},
we find the following inequality for very large $\omega$ values
\begin{equation}
\frac{1}{ a_0^2}\mathcal{K}^{2}+\frac{16\pi}{3}\rho_{0}>0,
\end{equation}
{which is always satisfied because of $\rho_{0}>0$}. Therefore, for {very large $\omega$ values,} there is a stability against the tensor perturbations
in the framework of Horava-Lifshitz gravity inspired by entropic gravity. For an arbitrary and not so large $\omega$ values,
the inequality (\ref{e4-5}) indicates an interplay between the cosmological parameters $a_0$, $\omega$, $\mathcal{K}$, $k$ and
$\rho_0$ of the Einstein static universe.
\section{Conclusion}
The existence of
the big bang singularity in the early universe is one of the essential problems in standard cosmology. To solve this problem
in the framework of GR, the so called "emergent universe scenario" \cite{Ellis} was introduced as an inflationary
cosmology without the inial singularity. Based on this scenario, the early universe before the transition
to the inflationary phase, has an initial state, known as the Einstein static state. In the framework of this cosmological model,
 the universe has emerged from an Einstein static state rather than a big bang singularity.
Clearly,   variety of the perturbations  in the early universe can affect
the stability of the initial
static state. On the other hand, the classical GR is not a appropriate theory at high energy states so that early universe is highly
influenced by various physical conditions which may result in some modifications of GR. By this motivation,  we have investigated the stability of initial Einstein
static state against the linear homogeneous scalar, vector and tensor perturbations
within the framework of the new class of UV
complete theory of gravity known as the deformed HL gravity inspired by entropic force scenario. It is shown that  there is no stable Einstein static universe against the  linear scalar perturbations, for the case of flat universe, $k=0$.   In the case of large values of dynamical parameter of
HL gravity $\omega$, for a positively curved universe $k>0$, the domination of the quintessence and phantom
matter fields with the barotropic equation of state parameter $\beta<-\frac{1}{3}$ is necessary while
for a negatively curved universe $k<0$,
the  matter fields  with $\beta>-\frac{1}{3}$ are needed to be the dominant fields of the universe.
There is a neutral stability against the vector perturbations.
For the tensor perturbations, an inequality  including the parameters $a_0$, $\omega$, $\mathcal{K}$, $k$ and
$\rho_0$ of the Einstein static universe is obtained which certainly accounts for a  stability, for large $\omega$ values.
 \section*{Acknowledgment}
This work has been supported financially by Research Institute
for Astronomy and Astrophysics of Maragha (RIAAM) under research project
NO.1/3252-39.

\end{document}